\documentclass{PoS}
\usepackage{amsmath}

\newcommand{\al}{\ensuremath{\alpha} }
\newcommand{\be}{\ensuremath{\beta} }
\newcommand{\ga}{\ensuremath{\gamma} }
\newcommand{\de}{\ensuremath{\delta} }

\newcommand{\eps}{\ensuremath{\epsilon} }
\newcommand{\la}{\ensuremath{\lambda} }
\newcommand{\lalat}{\ensuremath{\la_{\text{lat}}} }
\newcommand{\mulat}{\ensuremath{\mu_{\text{lat}}} }
\newcommand{\Irm}{\ensuremath{\textrm I} }
\newcommand{\Jrm}{\ensuremath{\textrm J} }
\newcommand{\Krm}{\ensuremath{\textrm K} }
\newcommand{\cD}{\ensuremath{\mathcal D} }
\newcommand{\cO}{\ensuremath{\mathcal O} }

\newcommand{\SO}[1]{\ensuremath{\text{SO(}#1\text{)}} }
\newcommand{\lra}{\ensuremath{\longrightarrow} }
\newcommand{\X}{\ensuremath{\!\times\!} }
\newcommand{\Tr}[1]{\ensuremath{\mbox{Tr}\left[ #1 \right]} }
\newcommand{\vev}[1]{\ensuremath{\left\langle #1 \right\rangle} }
\newcommand{\eq}[1]{Eq.~\ref{#1}}
\newcommand{\fig}[1]{Fig.~\ref{#1}}
\newcommand{\secref}[1]{Section~\ref{#1}}
\newcommand{\refcite}[1]{Ref.~\cite{#1}}
\newcommand{\mysection}[1]{\vspace{-6 pt}\section{#1}\vspace{-6 pt}}

\title{Thermal phase structure of a supersymmetric matrix model}
\ShortTitle{Thermal phase structure of a supersymmetric matrix model}
\author{\speaker{David Schaich} \\
  Department of Mathematical Sciences, University of Liverpool, \\ Liverpool L69 7ZL, United Kingdom \\
  E-mail: \email{david.schaich@liverpool.ac.uk}
}
\author{Raghav G.~Jha \\
  Perimeter Institute for Theoretical Physics, Waterloo, Ontario N2L 2Y5, Canada \\
  E-mail: \email{rjha1@perimeterinstitute.ca}
}
\author{Anosh Joseph \\
  Department of Physical Sciences, Indian Institute of Science Education and Research - Mohali, \\ Knowledge City, Sector 81, SAS Nagar, Punjab 140306, India \\
  E-mail: \email{anoshjoseph@iisermohali.ac.in}
}

\abstract{ 
  We present initial results from ongoing lattice investigations into the thermal phase structure of the Berenstein--Maldacena--Nastase deformation of maximally supersymmetric Yang--Mills quantum mechanics.
  The phase diagram of the theory depends on both the temperature $T$ and the deformation parameter $\mu$, through the dimensionless ratios $T / \mu$ and $g \equiv \la / \mu^3$ with \la the 't~Hooft coupling.
  Considering couplings $g$ that span three orders of magnitude, we reproduce the weak-coupling perturbative prediction for the deconfinement $T / \mu$ and approach recent large-$N$ dual supergravity analyses in the strong-coupling limit.
  We are carrying out calculations with lattice sizes up to $N_{\tau} = 24$ and numbers of colors up to $N = 16$, to allow initial checks of the large-$N$ continuum limit.
}

\FullConference{37th International Symposium on Lattice Field Theory \\ 16--22 June 2019 \\ Wuhan, China}

\begin{document}
\setlength{\abovedisplayskip}{6 pt}
\setlength{\belowdisplayskip}{6 pt}
\setcounter{section}{-1}
\mysection{Introduction, motivation and background} 
Supersymmetric matrix models in 0+1 dimensions, corresponding to dimensional reductions of supersymmetric Yang--Mills (SYM) theories, are compelling targets for current lattice calculations.
These systems involve balanced collections of interacting bosonic and fermionic $N\X N$ matrices at a single spatial point.
As recently reviewed by Refs.~\cite{Kadoh:2016eju, Schaich:2018mmv}, this setting dramatically reduces the difficulties inherent in investigating supersymmetry in a discretized space-time.
In particular, these theories tend to be super-renormalizable and in many cases a one-loop counterterm calculation suffices to restore supersymmetry in the continuum limit, with no need for the numerical fine-tuning typically required in higher dimensions.

In addition, maximally supersymmetric matrix models with $Q = 16$ supercharges retain conjectured `holographic' connections to dual quantum gravity theories.
This is exemplified by the conjecture by Banks, Fischler, Shenker and Susskind (BFSS~\cite{Banks:1996vh}) that the large-$N$ limit of $Q = 16$ SYM quantum mechanics describes the strong-coupling (M-theory) limit of Type~IIA string theory in the infinite-momentum frame.
This has motivated extensive numerical investigations of this `BFSS model' over the past decade.
(See the reviews \cite{Kadoh:2016eju, Schaich:2018mmv} for comprehensive references.)

One challenge faced by these studies of the BFSS model is the lack of a well-defined thermal partition function, due to the need to integrate over a non-compact moduli space corresponding to the radiation of D0-branes.
This makes Monte Carlo analyses unstable for $N < \infty$~\cite{Catterall:2009xn, Hanada:2009hq}; as $N$ increases the system may spend longer fluctuating around a metastable vacuum, but this will eventually decay. 
The standard cure for this instability is to add a scalar potential to the calculation, but this introduces an additional source of supersymmetry breaking, and an additional extrapolation to vanishing scalar potential that can significantly increase the complexity and costs of the calculations.

A more compelling alternative is to study the Berenstein--Maldacena--Nastase (BMN~\cite{Berenstein:2002jq}) deformation of the BFSS model, which describes the discrete light-cone quantization compactification of M-theory on the maximally supersymmetric ``pp-wave'' background of 11d supergravity (involving plane-fronted waves with parallel rays).
This deformation introduces non-zero masses for the nine scalars and sixteen fermions of the theory, explicitly breaking an SO(9) global symmetry (corresponding to the compactified spatial directions of ten-dimensional SYM) down to $\SO{6}\X \SO{3}$.
The resulting `BMN model' preserves all $Q = 16$ supercharges and has a well-defined thermal partition function, making it an attractive system to study using Monte Carlo methods.
Indeed, some lattice studies of the BMN model have already been carried out~\cite{Catterall:2010gf, Asano:2018nol}. 
There was also a recent lattice investigation of the non-supersymmetric `bosonic BMN model' that omits the fermionic degrees of freedom~\cite{Asano:2020yry}.

In this proceedings we report initial results from our ongoing investigations of the thermal phase structure of the BMN model.
We begin in the next section by briefly reviewing the model and our lattice discretization of it.
In \secref{sec:data} we discuss the lattice observables used to explore the phase diagram and present preliminary results for the critical $T / \mu$ of deconfinement over a wide range of couplings $g$.
We conclude in \secref{sec:conc} with a brief summary of the next steps remaining to complete this work.

\mysection{\label{sec:lattice}BMN model and lattice discretization} 
We begin with the dimensional reduction of ten-dimensional SYM with gauge group SU($N$), compactifying all nine spatial directions.
In 0+1~dimensions, the corresponding nine spatial components of the original ten-dimensional gauge field become the scalars $X_i$ with $i = 1, \cdots, 9$.
These scalars, the sixteen fermions $\Psi_{\al}$ with $\al = 1, \cdots, 16$ and the temporal component of the gauge field (in the covariant derivative $D_{\tau}$) are all $N\X N$ matrices evolving in time and transforming in the adjoint representation of SU($N$).
This defines the BFSS action, which we will call $S_0$, and to which we add the $\mu$-dependent BMN deformation $S_{\mu}$ that changes the dual geometry from flat space to the plane-wave geometry while preserving all $Q = 16$ supersymmetries.
With the gauge group generators normalized as $\Tr{T^A T^B} = -\de_{AB}$, the action is
\begin{align}
  S       & = S_0 + S_{\mu}                                                                 \label{eq:action} \\
  S_0     & = \frac{N}{4\la} \int d\tau \ \mbox{Tr} \Bigg[ -\left(D_{\tau} X_i\right)^2 + \Psi_{\al}^T \ga_{\al\be}^{\tau} D_{\tau} \Psi_{\be}  - \frac{1}{2} \sum_{i < j} \left[X_i, X_j\right]^2 + \frac{1}{\sqrt 2} \Psi_{\al}^T \ga_{\al\be}^i \left[X_i, \Psi_{\be}\right]\Bigg] \cr
  S_{\mu} & = -\frac{N}{4\la} \int d\tau \ \mbox{Tr} \Bigg[ \left(\frac{\mu}{3} X_{\Irm}\right)^2 + \left(\frac{\mu}{6} X_A\right)^2 + \frac{\mu}{4} \Psi_{\al}^T \ga_{\al \be}^{123} \Psi_{\be} - \frac{\sqrt{2} \mu}{3} \eps_{\Irm\Jrm\Krm} X_{\Irm} X_{\Jrm} X_{\Krm}\Bigg].       \nonumber
\end{align}
In this (0+1)-dimensional setting, the 't~Hooft coupling $\la \equiv g_{\text{YM}}^2 N$ is dimensionful.
$\ga^{\tau}$ and $\ga^i$ are the $16\X 16$ Euclidean gamma matrices, with $\ga^{123} \equiv \frac{1}{3!} \eps_{\Irm\Jrm\Krm} \ga^{\Irm} \ga^{\Jrm} \ga^{\Krm}$.
We use a representation where
\begin{equation*}
  \ga^{\tau} = \begin{pmatrix}0   & I_8 \\
                              I_8 & 0\end{pmatrix}.
\end{equation*}

In the final line of \eq{eq:action}, the indices $\Irm, \Jrm, \Krm = 1, 2, 3$ while $A = 4, \cdots 9$, splitting the nine scalars into two sets: an SO(6) multiplet $X_A$ and an SO(3) multiplet $X_{\Irm}$ with a different mass and with a trilinear interaction known as the Myers term.
This produces the explicit $\SO{9} \lra \SO{6}\X \SO{3}$ global symmetry breaking mentioned in the introduction.

In the interest of simplicity, for the time being we are working with the most direct lattice discretization of this system.
With thermal boundary conditions (periodic for the bosons and anti-periodic for the fermions), we replace the temporal extent of the system with $\be = aN_{\tau}$ corresponding to $N_{\tau}$ sites separated by lattice spacing `$a$'.
This defines the dimensionless temperature $t = 1 / N_{\tau}$ with $T = t / a$. 
The dimensionless lattice parameters $\mulat = a\mu$ and $\lalat = a^3\la$ are combined just like the corresponding continuum quantities to define $g = \lalat / \mulat^3$, while $T / \mu = 1 / (N_{\tau} \mulat)$.

We use the following prescription for the lattice finite-difference operator that replaces the covariant derivative:
\begin{equation}
  \cD_{\tau} = \begin{pmatrix}0            & \cD_{\tau}^+ \\
                              \cD_{\tau}^- & 0\end{pmatrix},
\end{equation}
where $\cD_{\tau}^-$ is the adjoint of $\cD_{\tau}^+$, which acts on the fermions at lattice site $n$ as
\begin{equation}
  \cD_{\tau}^+ \Psi_n = U_{\tau}(n) \Psi_{n + 1} U_{\tau}(n)^{\dag} - \Psi_n.
\end{equation}
Here $U_{\tau}(n)$ is the Wilson gauge link connecting site $n$ with site $n + 1$.
$U_{\tau}^{\dag}$ is the adjoint link with the opposite orientation.
This nearest-neighbor finite-difference operator produces the correct number of fermion degrees of freedom, with no extraneous `doublers'. 

Our numerical calculations use the standard rational hybrid Monte Carlo algorithm, which we have implemented in a publicly available development branch of the parallel lattice supersymmetry software described by \refcite{Schaich:2014pda}.\footnote{\texttt{\href{https://github.com/daschaich/susy}{github.com/daschaich/susy}}}
The addition of the BMN model to this software package, along with related improvements to the large-$N$ performance of the code and other developments in higher dimensions, will soon be presented in another publication~\cite{parallel_imp}.

\begin{figure}[tbp]
  \includegraphics[width=0.45\linewidth]{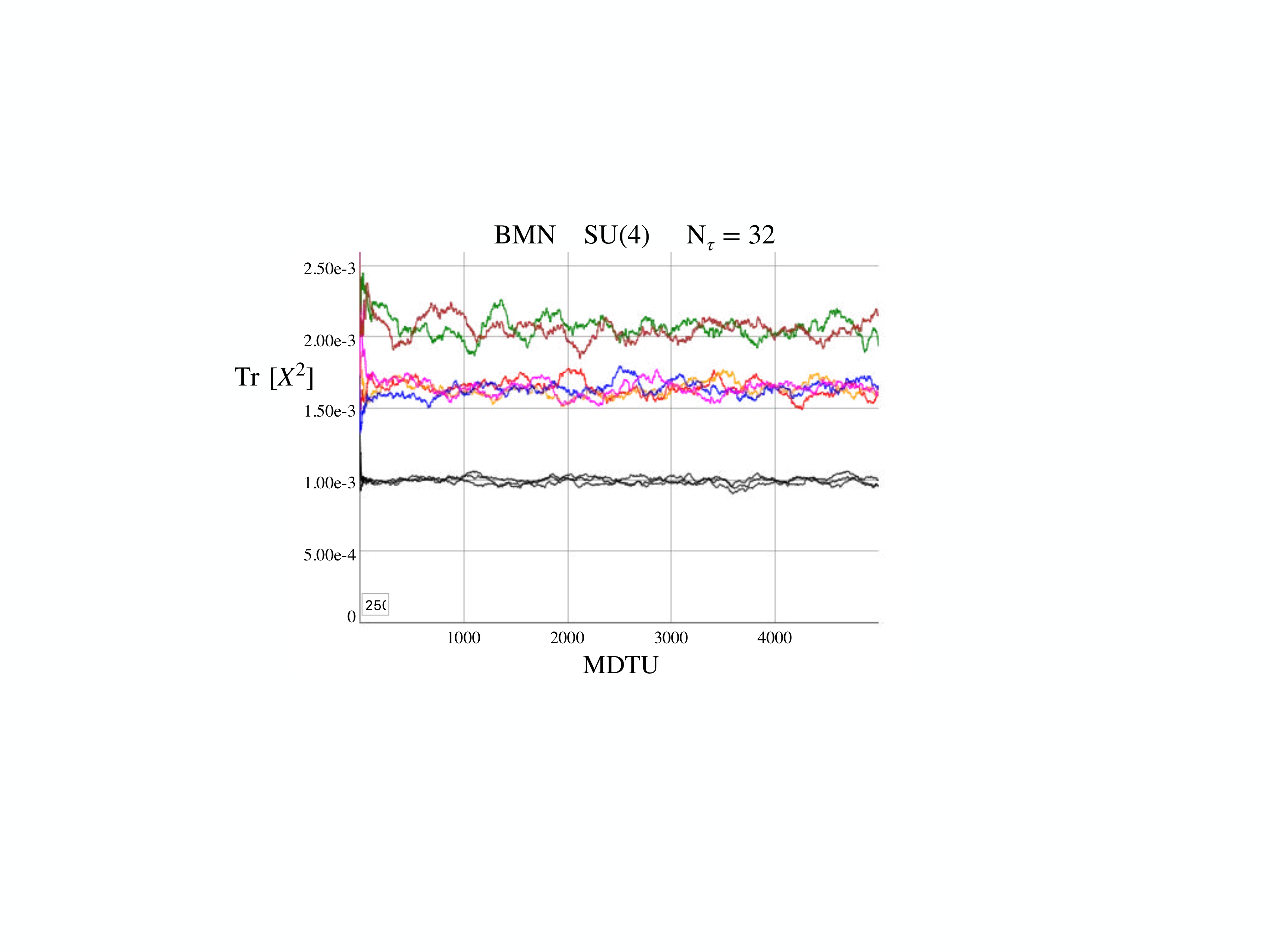}\hfill \includegraphics[width=0.45\linewidth]{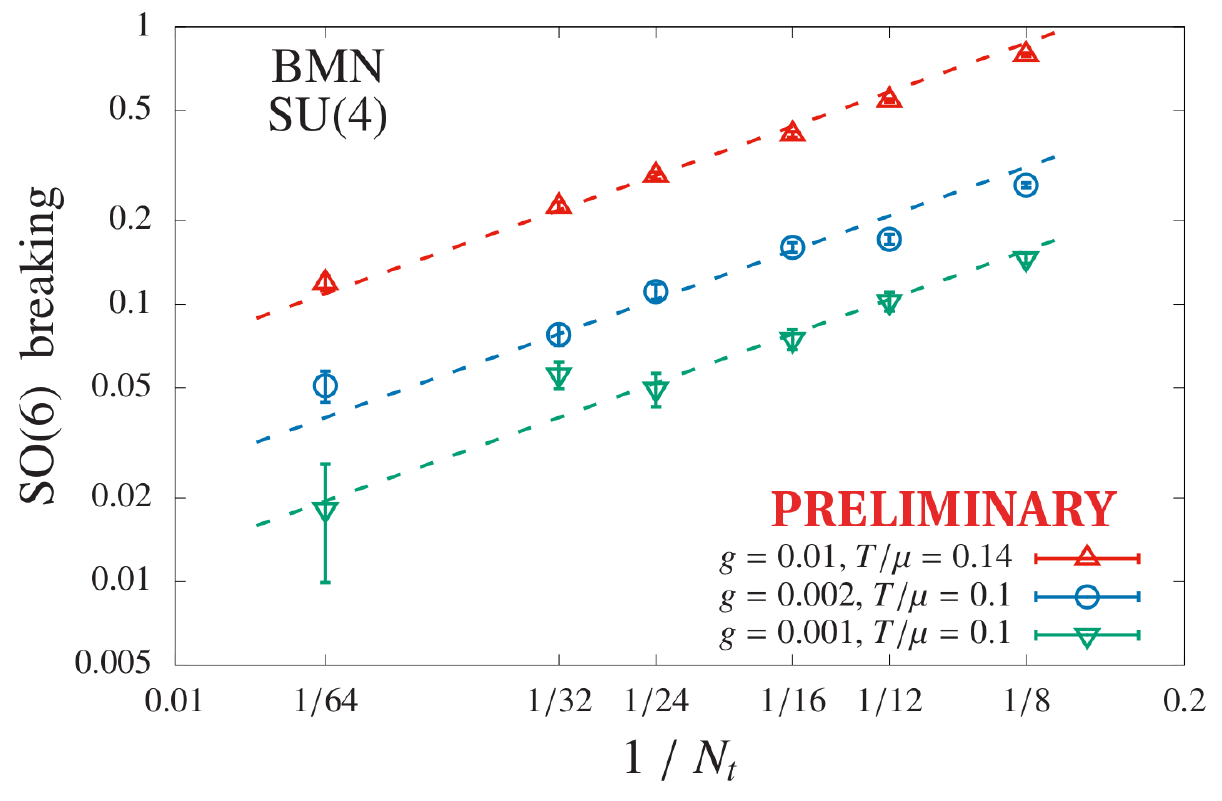}
  \caption{\label{fig:splitting}\textbf{Left:} Time-series plot of $\Tr{X^2}$ data for an SU(4) ensemble with $N_{\tau} = 32$ sites, $g = 0.01$ and $T / \mu = 0.14$, showing running averages over 250 molecular dynamics time units (MDTU).  The three black (six colored) lines at the bottom (top) of the plot illustrate the expected $\SO{9} \lra \SO{6}\X \SO{3}$ global symmetry breaking.  The further splitting of the six colored $\Tr{X_A^2}$ into groups of two and four indicates that the SO(6) symmetry is also broken.  \textbf{Right:} Confirmation that this SO(6) symmetry breaking is a discretization artifact, by showing that the breaking measure defined in \protect\eq{eq:splitting} vanishes in the $N_{\tau} \to \infty$ continuum limit.  The dashed lines are not fits, but rather illustrate the expected behavior of $\cO(a)$ discretization artifacts.}
\end{figure}

A drawback of using such a simple lattice action is that discretization artifacts can be significant for small $N_{\tau}$, especially as the dimensionless coupling $g$ increases.
This is illustrated in \fig{fig:splitting}, where the discretization artifact in question is a breakdown of the expected SO(6) symmetry for the gauge-invariant observable $\Tr{X_A^2}$.
The left plot shows these six scalar-square traces splitting into groups of two and four.
In the right plot we quantify this splitting by computing the ratio
\begin{equation}
  \label{eq:splitting}
  \frac{\vev{\Tr{X_{(2)}^2}} - \vev{\Tr{X_{(4)}^2}}}{\vev{\Tr{X_{(6)}^2}}},
\end{equation}
where $\vev{\Tr{X_{(2)}^2}}$, $\vev{\Tr{X_{(4)}^2}}$ and $\vev{\Tr{X_{(6)}^2}}$ average over the two larger traces, the four smaller traces and all six of them, respectively.
The key feature that allows us to identify this SO(6) symmetry breaking as a discretization artifact is the observation that this ratio vanishes in the $N_{\tau} \to \infty$ continuum limit.
This is shown by the dashed lines in the plot, which are not fits but rather illustrate the expected behavior of $\cO(a)$ discretization artifacts. 

\mysection{\label{sec:data}Observables and preliminary results} 
Using the code discussed above, we are analyzing a range of couplings $10^{-5} \leq g \leq 10^{-2}$ for SU($N$) gauge groups with $N = 8$, 12 and 16 using lattice sizes $N_{\tau} = 8$, 16 and 24.
For each $\left\{N, N_{\tau}, g\right\}$, we scan in the dimensionless ratio $T / \mu$ to determine the critical value at which the theory transitions between confined and deconfined phases.
Our goal is to non-perturbatively connect the two limits that have been investigated analytically.
In the limit of strong coupling $g = \lalat / \mulat^3 \to \infty$ (with $T / \la^{1/3} = 1 / (N_{\tau} \lalat^{1 / 3}) \ll 1$), the supergravity calculations of \refcite{Costa:2014wya} predict that the first-order deconfinement transition occurs at the critical temperature
\begin{equation}
  \left.\lim_{g \to \infty} \frac{T}{\mu}\right|_{\textrm{crit}} = 0.105905(57).
\end{equation}
In the opposite limit of weak coupling $g \to 0$, perturbative calculations~\cite{Furuuchi:2003sy, Spradlin:2004sx, Hadizadeh:2004bf} predict a smaller critical temperature
\begin{align}
  \label{eq:NNLO}
  \left.\lim_{g \to 0} \frac{T}{\mu}\right|_{\textrm{crit}} & = \lim_{g \to 0} \frac{1}{12\log 3} \left(1 + \frac{2^6 \cdot 5}{3^4}\left(27 g\right) - C_{\text{NNLO}}\left(27 g\right)^2 + \cO\left(27^3 g^3\right)\right) \approx 0.076 \\
                                                            & C_{\text{NNLO}} = \frac{23\cdot 19\,927}{2^2\cdot 3^7} + \frac{1765769\log 3}{2^4\cdot 3^8}, \nonumber
\end{align}
with the transition again first order.
(The appropriate perturbative expansion parameter is $27g$.)

\begin{figure}[tbp]
  \includegraphics[width=0.45\linewidth]{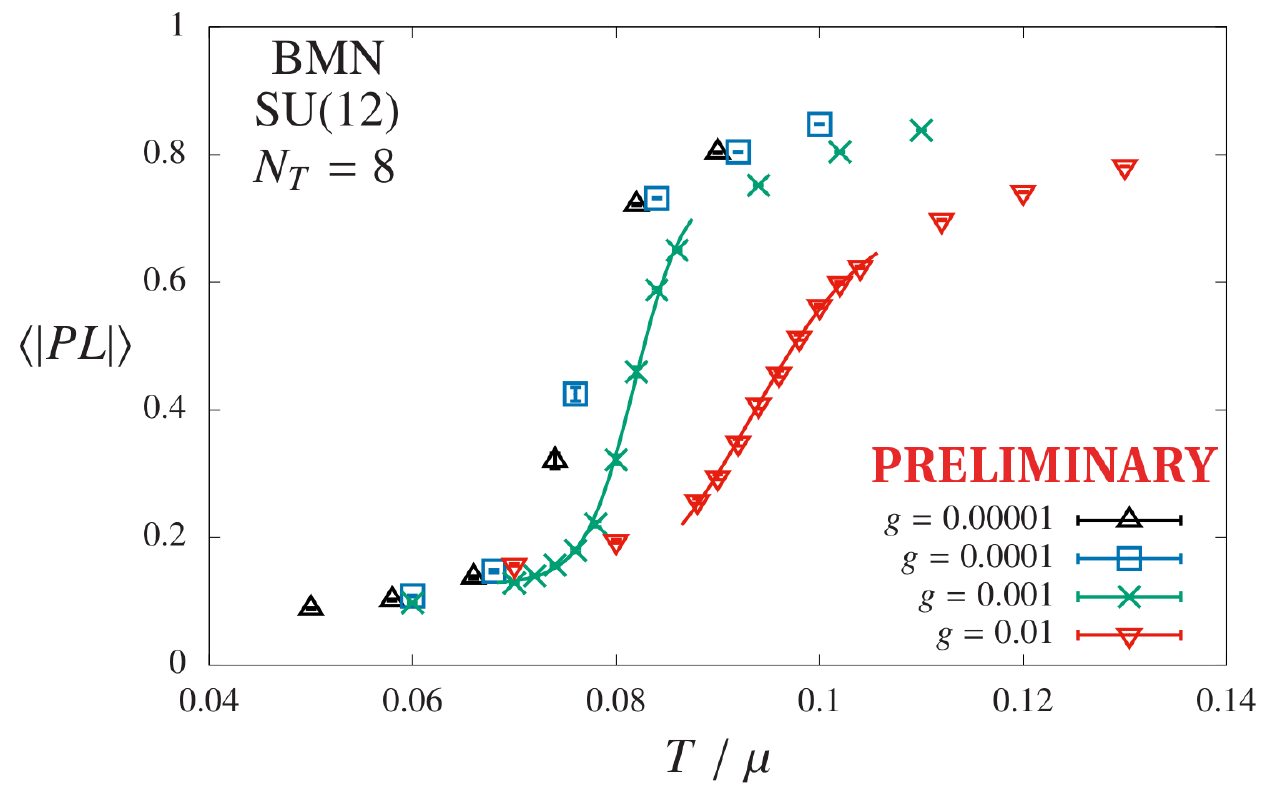}\hfill \includegraphics[width=0.45\linewidth]{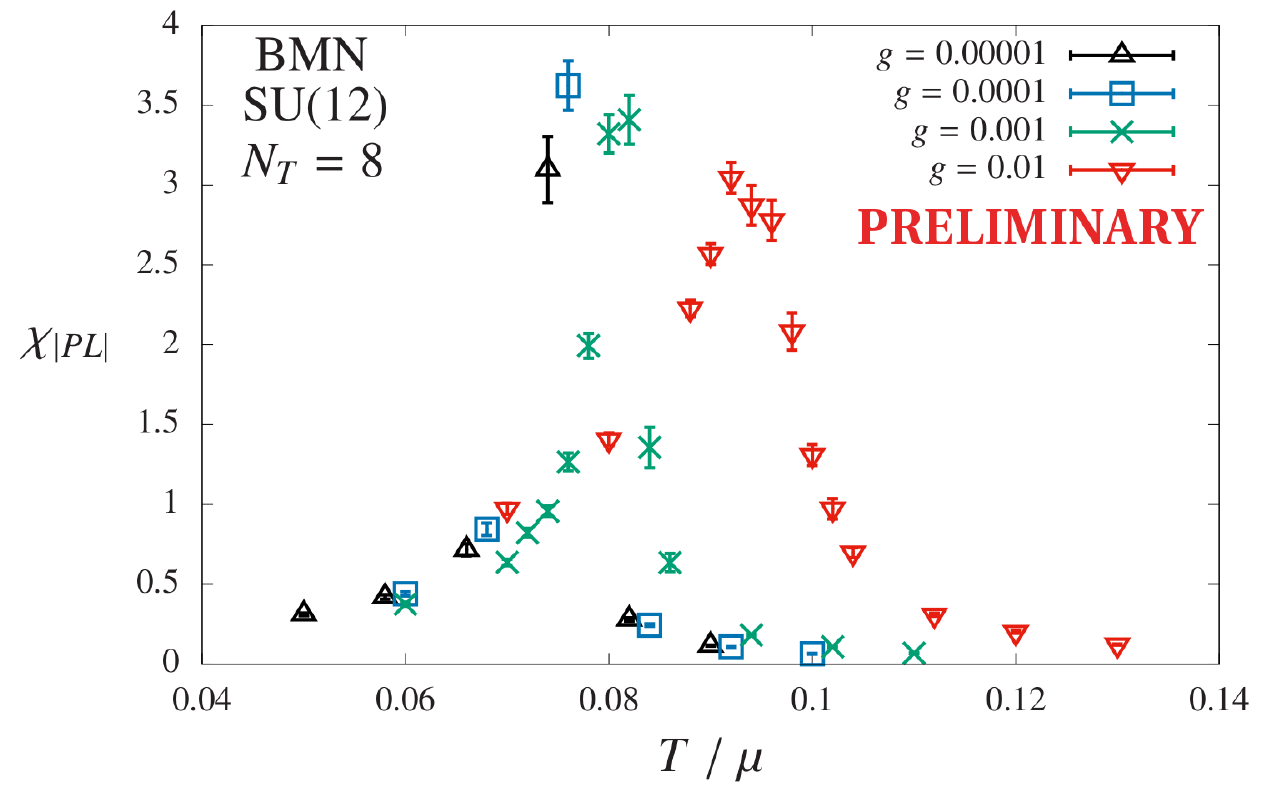}
  \caption{\label{fig:poly}Representative Polyakov loop data from which we can determine the critical $T / \mu$ of the deconfinement transition, for gauge group SU(12) with $N_{\tau} = 8$ lattice sites.  \textbf{Left:} We interpolate the Polyakov loop itself using \protect\eq{eq:sigmoid} to determine the critical $T / \mu$.  \textbf{Right:} We confirm that the Polyakov loop susceptibility shows peaks at the same critical $T / \mu$.}
\end{figure}

The main (but not only) observable we use to determine the critical $T / \mu$ of the transition is the Polyakov loop (i.e., the absolute value of the trace of the holonomy along the temporal circle).
Figure~\ref{fig:poly} presents some representative results for the Polyakov loop for a subset of our ongoing calculations with gauge group SU(12) and $N_{\tau} = 8$ lattice sites, considering four values of the coupling $g = 10^{-5}$, $10^{-4}$, $10^{-3}$ and $10^{-2}$.
The left plot shows the Polyakov loop itself, along with a couple of interpolations using a simple four-parameter sigmoid ansatz,
\begin{equation}
  \label{eq:sigmoid}
  P = A - \frac{B}{1 + \exp\left[C (T / \mu - D)\right]}.
\end{equation}
From these interpolations we can read off the critical $T / \mu$ as the fit parameter $D$ that corresponds to the location of the maximum slope of the sigmoid.
In the right plot we confirm that the Polyakov loop susceptibilities have peaks at the same critical values of $T / \mu$.
In addition to the SU(8) and SU(16) calculations we don't include in this proceedings, more SU(12) runs are currently underway to fill in a higher density of points in these plots, in particular for smaller $g \leq 10^{-4}$.

\begin{figure}[tbp]
  \centering
  \includegraphics[width=0.7\linewidth]{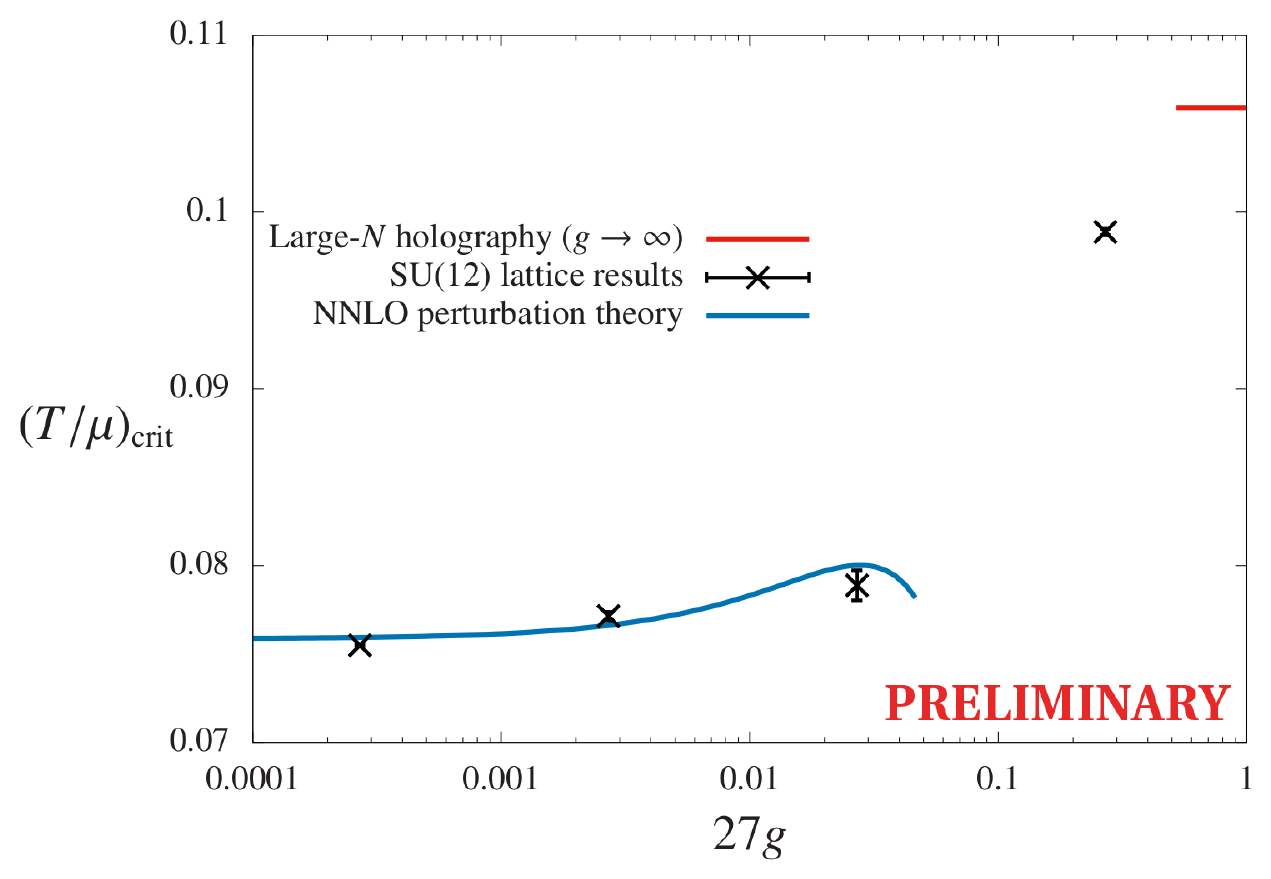}\hfill
  \caption{\label{fig:phase_diag}Preliminary lattice results for the BMN phase diagram in the plane of critical $T / \mu$ vs.\ the perturbative expansion parameter $27g$ (where $g$ is the dimensionless coupling), considering only SU(12) numerical computations.  The uncertainties on the data points are currently dominated by the $N_{\tau} \to \infty$ continuum extrapolations.  For sufficiently weak couplings the non-perturbative lattice results agree with the perturbative expression in \protect\eq{eq:NNLO}, shown as a blue curve.  As the coupling increases, they move towards the large-$N$ strong-coupling ($g \to \infty$) holographic prediction shown in red.}
\end{figure}

Figure~\ref{fig:phase_diag} presents our preliminary lattice results for the BMN phase diagram in the plane of critical $T / \mu$ vs.\ the perturbative expansion parameter $27g$, considering only the SU(12) analyses shown in \fig{fig:poly}.
We assign uncertainties to our results by carrying out initial $N_{\tau} \to \infty$ continuum extrapolations that will be improved as we accumulate additional data.
Upon the completion of the corresponding SU(8) and SU(16) calculations, we will also be able to extrapolate to the large-$N$ limit relevant to holography.
Already we can see good agreement with the next-to-next-to-leading-order perturbative expression in \eq{eq:NNLO} (the blue curve) for sufficiently weak couplings.
At stronger couplings perturbation theory breaks down, while our results move in the expected direction, towards the large-$N$ strong-coupling ($g \to \infty$) holographic prediction shown in red.

\mysection{\label{sec:conc}Outlook and next steps} 
The preliminary results presented in \fig{fig:phase_diag} provide encouraging indications that our non-perturbative lattice investigations of the phase diagram of the BMN model (mass-deformed SYM quantum mechanics) will be able to interpolate between the weak-coupling perturbative regime and the strong-coupling large-$N$ holographic regime.
Much more work obviously remains to be completed.
In addition to finalizing the numerical calculations with gauge group SU(12) shown above, which will provide improved $N_{\tau} \to \infty$ continuum extrapolations, additional SU(8) and SU(16) calculations are also underway in order to enable extrapolations to the large-$N$ limit.
We are also analyzing more observables than can be discussed here, including the distributions of Polyakov loop eigenvalues and the free energy of the dual black hole, the latter of which can be compared with expectations based on supergravity.
After completing these analyses, it may also be worthwhile to explore more elaborate lattice discretizations of the BMN model, with the aim of reducing the discretization artifacts discussed in \secref{sec:lattice}.


\vspace{20 pt} 
\noindent \textsc{Acknowledgments:}~We thank Simon Catterall and Toby Wiseman for ongoing collaboration on lattice supersymmetry, Masanori Hanada for interesting discussions, and Yuhma Asano and Denjoe O'Connor for helpful conversations and sharing numerical data from \refcite{Asano:2018nol}.
Numerical calculations were carried out at the University of Liverpool and at the San Diego Computing Center through XSEDE supported by National Science Foundation grant number ACI-1548562. 
Research at Perimeter Institute is supported in part by the Government of Canada through the Department of Innovation, Science and Economic Development Canada and by the Province of Ontario through the Ministry of Economic Development, Job Creation and Trade.
AJ was supported in part by a Start-up Research Grant (No.~{SRG/2019/002035}) from the Science and Engineering Research Board, Department of Science and Technology, Government of India; a Seed Grant from the Indian Institute of Science Education and Research Mohali; and by the European Research Council under the European Union's Seventh Framework Programme ({FP7/2007-2013}), ERC grant agreement {STG279943}, ``Strongly Coupled Systems''.
DS was supported by UK Research and Innovation Future Leader Fellowship {MR/S015418/1}.

\bibliographystyle{utphys}
\bibliography{lattice19}
\end{document}